\documentclass[prc,twocolumn,floatfix,superscriptaddress]{revtex4-2}
\usepackage[pdftex,plainpages=false,colorlinks=true,linkcolor=blue, citecolor=blue, urlcolor=blue]{hyperref}
\usepackage{amsfonts}
\usepackage{amsmath}
\usepackage{amssymb}
\usepackage{graphicx}
\usepackage{natbib}
\usepackage{color}
\usepackage{placeins}
\usepackage{physics}

\begin{document}
 
\title{Introducing the concept of the Widom line in the QCD phase diagram}
\author{G. Sordi}
\email[corresponding author: ]{giovanni.sordi@rhul.ac.uk}
\affiliation{Department of Physics, Royal Holloway, University of London, Egham, Surrey, UK, TW20 0EX}
\author{A.-M. S. Tremblay}
\affiliation{D\'epartement de physique, Institut quantique \& RQMP, Universit\'e de Sherbrooke, Sherbrooke, Qu\'ebec, Canada J1K 2R1}
\date{\today}

\begin{abstract}
Critical phenomena emerging from the critical end point of a first-order transition are ubiquitous in nature. Here we bring the concept of a supercritical crossover, the Widom line, initially developed in the context of fluids, into the interacting matter described by quantum chromodynamics (QCD). We show that the existence of the putative critical end point between hadron gas and quark-gluon plasma in the temperature versus chemical potential of the QCD phase diagram implies the existence of a Widom line emerging from it in the supercritical region. We survey the thermodynamic anomalies already identified in simplified theoretical models of QCD exhibiting a critical end point, to show that they can be interpreted in terms of a Widom line. Then we suggest possible directions where the Widom line concept could provide new light on the QCD phase diagram.  
\end{abstract}
 
\maketitle

\section{Introduction}
Understanding whether the phase diagram of quantum chromodynamics (QCD) hosts a critical end point between a gas of hadrons and a quark-gluon plasma remains a central challenge~\cite{StephanovRev, KogutStephanov:QCDBook, QCDrev4-2022, QCDrev1-2023, QCDrev2-2023, QCDrev3-2023}. Intense theoretical and experimental efforts are currently devoted to unravel this issue~\cite{QCDrev4-2022, StephanovRev}. 

From a theory perspective, our current understanding of the behavior of a system close to a critical point is shaped by the ideas of the renormalization group approach. In this approach, the concept of universality -- quite generally, the independence of some macroscopic properties from the microscopic details of a system -- is of key importance~\cite{stanley1999scaling}. For example, there is a deep analogy between the phase transition between liquid and gas and the transition between a ferromagnet and a paramagnet. 

As a consequence, the concept of universality has often led to valuable exchange of ideas between different fields. 
In this spirit, here we shall bring the conceptual framework of the Widom line, developed in condensed matter, to the interacting matter described by QCD. The Widom line is a supercritical crossover emerging from the critical end point of a first-order transition and serves to quantify the proximity to the critical region. 
This concept, developed by Eugene Stanley and coworkers in 2005~\cite{water1}, offers a novel perspective on the supercritical region of the critical end point of a first-order transition and has since been widely used in condensed matter, especially for the description of crossovers in classical fluids and electronic fluids, but not yet in QCD. 

Therefore, the main aims of this paper are (i) to introduce the conceptual framework of the Widom line in QCD matter, (ii) to show that the thermodynamic anomalies identified in prior work using simplified mathematical models of QCD can be interpreted within the  ``Widom line'' framework, and (iii) to suggest some directions where the concept of Widom line has the potential to bring new perspectives on the QCD phase diagram. 

\section{The concept of Widom line}
The Widom line, named after Benjamin Widom, has been introduced in Ref.~\cite{water1} in the context of fluids. It is a supercritical crossover emerging from the critical end point of a first-order transition. It is defined as the locus of maximum correlation length $\xi$ in the supercritical region~\cite{water1, supercritical, Franzese_JPCM2007}. At the critical end point the correlation length $\xi$ diverges. In the language of the renormalization group, the Widom line is the line of zero ordering field~\cite{Luo_PRL2014}. We will return on this formal definition of the Widom line in Sec.~\ref{Sec:models}.

Operationally, the Widom line can be approximated by the locus of local extrema of thermodynamic response functions~\cite{water1, supercritical, Luo_PRL2014}. This is because asymptotically close to the end point, all response functions are proportional to powers of the correlation length $\xi$, so the extrema of the response functions converge into one line -- the Widom line -- asymptotically close the critical end point~\cite{water1, Franzese_JPCM2007, Luo_PRL2014}. Note that the extrema of thermodynamic response functions have distinctive features: they are broad and weak in magnitude far from the end point and become narrow and pronounced in magnitude upon approaching the end point, where they diverge. These distinctive features, controlled by the Widom line, can be used to measure the proximity to the critical end point. 

\begin{figure}[ht!]
\centering{
\includegraphics[width=0.999\linewidth]{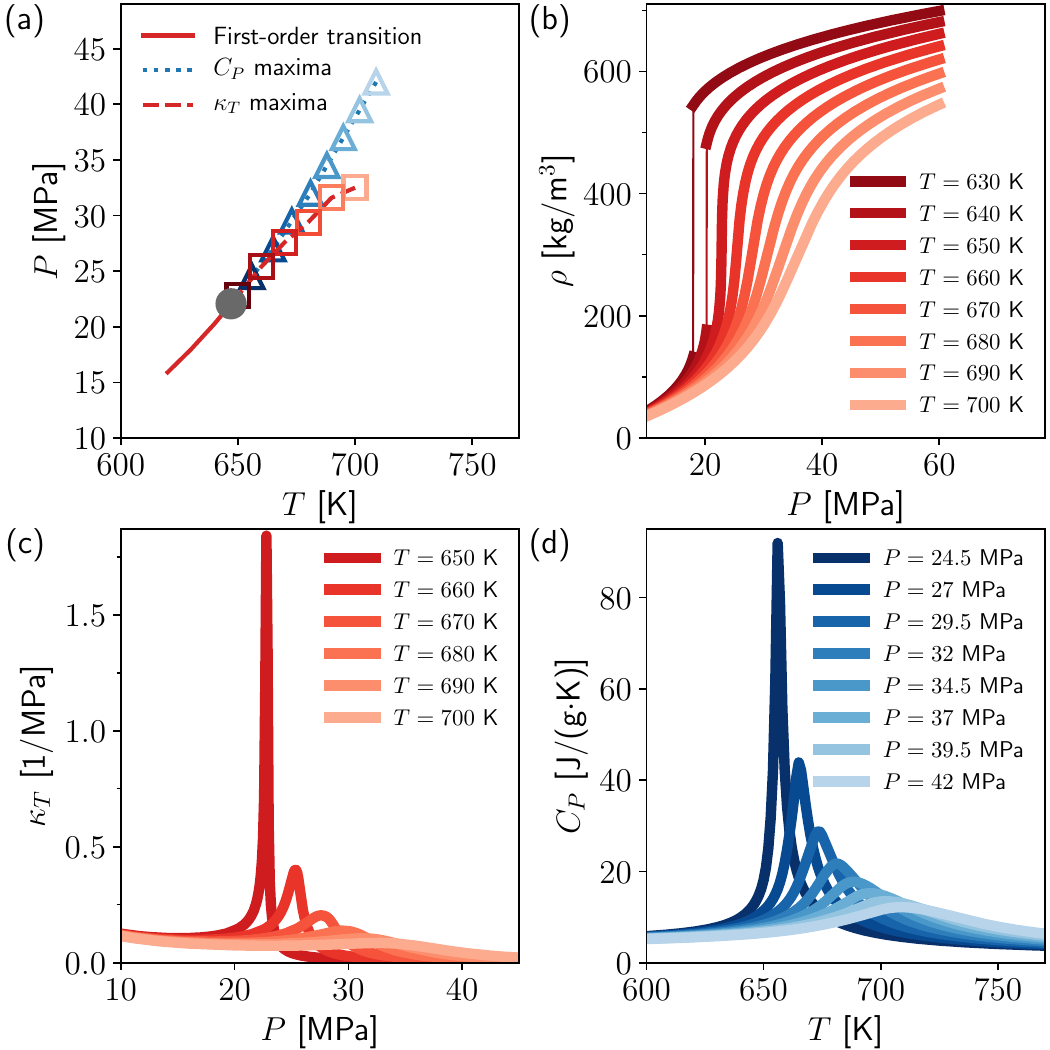}
}
\caption{Widom line concept for the liquid-gas transition in water. Experimental data are taken from Ref.~\cite{NISTWebBook}, following the procedure described in Ref.~\cite{Corradini_NatComm2014}. (a) Pressure $P$ vs temperature $T$ phase diagram of water. Solid red line indicates the liquid-gas first-order transition, terminating in the critical point at $(T_c, P_c)$ (full gray circle). From the critical end point crossovers emerge. They are formed by the locus of local extrema of thermodynamic response functions. These crossovers approximate the Widom line. (b) Density $\rho$ vs pressure $P$ for different values of temperature.  Below $T_c$, $\rho(P)$ is discontinuous. Above $T_c$, $\rho(P)$ is continuous and displays an inflection point. (c) Isothermal compressibility $\kappa_T(P)=\rho^{-1} (d\rho/dP)_T$ for different values of $T$ above $T_c$. The loci of $\kappa_T$ maxima are denoted by open red squares in panel (a). (d) Isobaric heat capacity $C_P(T)$ for different values of $P$ above $P_c$. The loci of $C_P$ maxima are indicated by open blue triangles in panel (a).
}
\label{fig:water}
\end{figure}

For simplicity, let us exemplify the Widom line concept by considering the familiar liquid-gas transition in water (see Figure~\ref{fig:water}). It has been shown~\cite{Corradini_NatComm2014, Gallo_ChemRev2016} that in the pressure $P$ versus temperature $T$ phase diagram, thermodynamic response functions such as isothermal compressibility $\kappa_T$ (Figure~\ref{fig:water}(c)), isobaric heat capacity $C_P$ (Figure~\ref{fig:water}(d)), isobaric expansion coefficient $\alpha_P$ all show maxima upon crossing the Widom line, and eventually diverge at the liquid-gas critical end point. 

The Widom line has been originally introduced in the context of supercooled water~\cite{water1, supercritical}, and has been discussed in the context of several other fluids~\cite{water1, supercritical, Franzese_JPCM2007}. It was then extended to a completely different state of matter, the strongly correlated electronic fluid obtained by doping a correlated insulator (Mott insulator) in Ref.~\cite{ssht}, suggesting its generality, as proposed in Ref.~\cite{Lorenzo3band}. It has since then been applied to different strongly correlated electronic systems.

The concept of the Widom line is important for two main reasons. 
First, by rationalising the behavior of different thermodynamic response functions, the Widom line concept provides a unifying theoretical framework to characterise supercritical crossovers above the critical end point of a first-order transition. This characterisation opens up novel theoretical predictions and may even unveil a novel microscopic understanding of crossovers. 

Second, by measuring the proximity to the critical region, it is an ``early predictor" of a critical end point, i.e. it allows to infer the location of a critical end point even when this critical end point is in a region inaccessible to experiments or is hidden by another phase. Let us discuss two important examples. In supercooled water~\cite{water1, supercritical}, the Widom line concept has been introduced as an indicator to predict the existence of a putative liquid-liquid transition hidden in the so called ``no man's land", a region inaccessible to experiments due to ice crystallisation. For electronic fluids, the concept of Widom line has been developed to predict the existence of a putative metal-metal transition hidden below the superconducting phase of hole-doped cuprates~\cite{ssht, sshtSC}.

\section{Widom line in the QCD phase diagram}
\label{Sec:WL-QCD}

\begin{figure}[ht!]
\centering{
\includegraphics[width=0.985\linewidth]{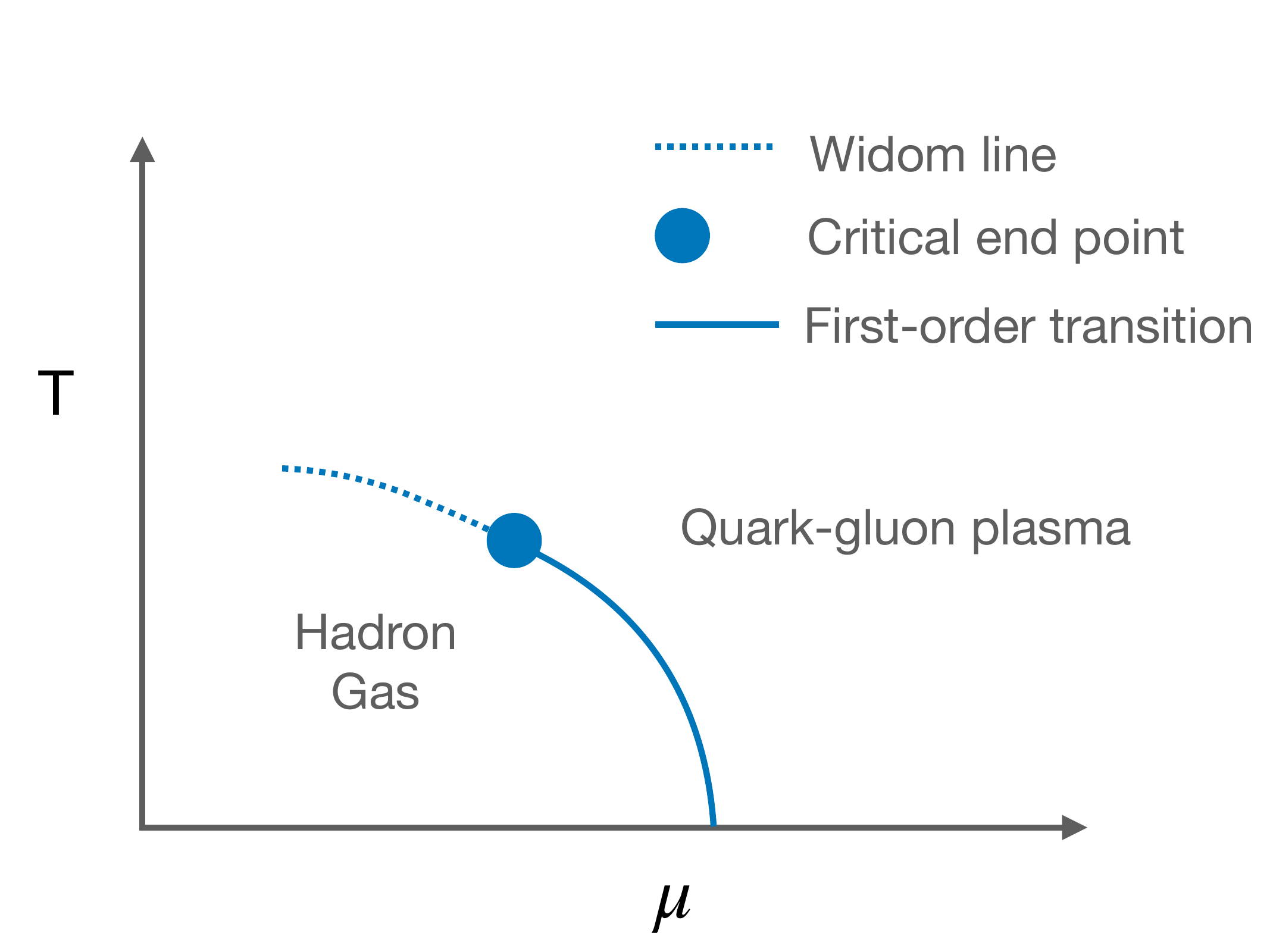}
}
\caption{Sketch of the QCD phase diagram according to the hypothesis of a critical end point at finite $T$ and $\mu$. The Widom line (dotted line) is the supercritical crossover emanating from the critical end point (full circle) of a first-order transition (solid line) between hadron gas and quark-gluon plasma.
}
\label{phasediagram}
\end{figure}

As a first step, in this section we apply the Widom line concept to the strongly correlated fluid realised in the QCD phase diagram. 
As sketched in Figure~\ref{phasediagram}, in the temperature $T$ versus baryon chemical potential $\mu$ phase diagram of QCD, it has been speculated that there lies a first-order transition between hadronic matter and a quark gluon plasma (continuous line). This transition ends at a second order critical point (full circle) at finite temperature and finite chemical potential. Based on physical grounds, the critical point is supposed to belong to the universality class of the 3D Ising model~\cite{Rajagopal_NuclPhysB1993, Berges_NuclPhysB1999, Karsch_PhysLettB2001, DeForcrand_NuclPhysB2003} (see Ref.~\onlinecite{StephanovRev} for a review of the QCD phase diagram).

The Widom line concept is a generic phenomenon appearing in the supercritical region of any finite temperature critical end point of a first-order transition. Hence, if the critical end point of the hadronic matter to quark gluon plasma transition exists, then the Widom line should emerge from the critical end point and extend in the supercritical region (dotted line in Figure~\ref{phasediagram}). 
The converse is also useful to consider: the Widom line and its manifestation as extrema of response functions becoming closer to each other can be used to support the existence of a critical end point. This is the strategy used for the conjectured liquid-liquid transition in supercooled water~\cite{water1, Franzese_JPCM2007} or for the conjectured metal-metal transition below the superconducting phase in hole-doped cuprates~\cite{ssht, sshtSC, Lorenzo3band}. 

The introduction of the Widom line in the QCD phase diagram is the first key contribution of this work. 

Note that in the QCD phase diagram there can exist multiple end points and hence multiple Widom lines. We discussed the hadronic matter to quark-gluon plasma critical point, but at small $T$ and small $\mu$ there is a nuclear liquid-gas end point~\cite{NuclearCP:1995} (not shown in Figure~\ref{phasediagram}). Similarly, in the phase diagram of water there is the familiar liquid-gas critical end point and the putative liquid-liquid critical end point, each of which has an associated Widom line. 

\section{Signs of Widom line in theoretical models of QCD}
\label{Sec:models}
\begin{figure}[ht!]
\centering{
\includegraphics[width=0.999\linewidth]{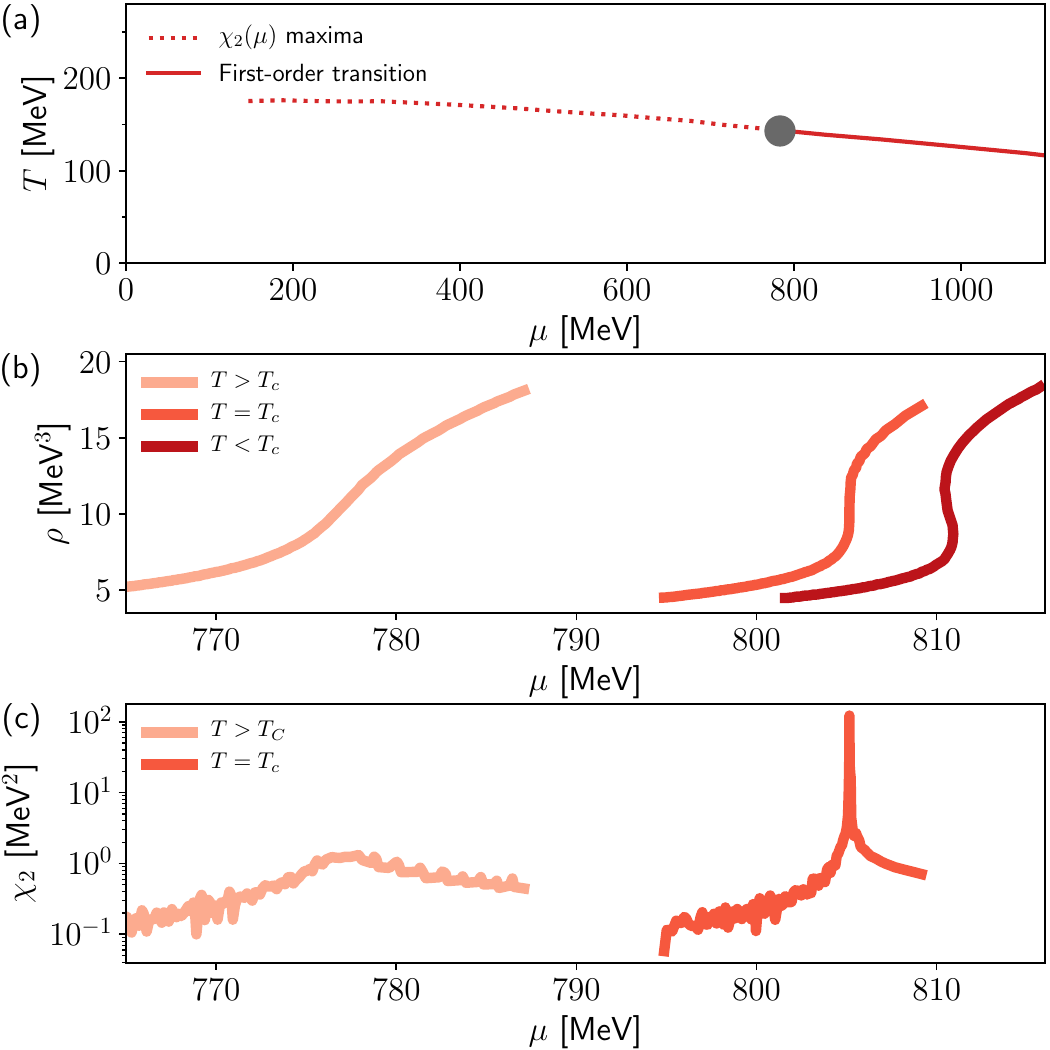}
}
\caption{QCD thermodynamics based on the Einstein-Maxwell-dilaton holographic model of Ref.~\cite{DeWolfe_PRD2011}. (a) Sketch of the QCD phase diagram. Full gray circle denotes the QCD critical end point, which in Ref.~\cite{DeWolfe_PRD2011} has been estimated to occur at $\mu_c=783$~MeV and $T_c=143$~MeV. Lines are guides to the eye: solid line indicates the first-order transition, and dotted line indicates the locus of the maxima of the quark susceptibility $\chi_2=(d\rho/d\mu)_T$, which is a proxy of the Widom line. (b) Baryon density $\rho$ as a function of chemical potential $\mu$ for different values of temperature $T$. Inflections develop in the supercritical region. Data are extracted from Ref.~\cite{DeWolfe_PRD2011}. Due to the extraction technique, the curves capture only the qualitative trends of the original data of Ref.~\cite{DeWolfe_PRD2011}. (c) Quark susceptibility $\chi_2=(d\rho/d\mu)_T$ as obtained by numerical derivative of the data in panel (b). A maximum appear: it is broad and weak in magnitude far from the critical end point and becomes narrow and sharp in magnitude near the critical end point, where it diverges. 
}
\label{fig:EDmodel}
\end{figure}

In this section we survey some thermodynamic anomalies reported in prior work on theoretical models of QCD which exhibit a critical end point at finite $T$ and finite $\mu$. The generality of the concept of Widom line implies that these models must exhibit a Widom line emerging from the critical end point. We shall show that this is indeed the case. We review two such models, without claim of completeness, but just to show the usefulness of the Widom line concept.
 
It is important to stress that critical fluctuations emerging from the QCD end point have long been expected to generate thermodynamic anomalies~\cite{StephanovRev}, and that the thermodynamic anomalies detected in theoretical models of QCD that contain a critical end point have also been already ascribed to the presence of the critical end point. Our contribution in this section is simply to rationalise the thermodynamic anomalies obtained in previous work within the conceptual framework of the ``Widom line". 

A first example deals with the Einstein-Maxwell-dilaton holographic model~\cite{DeWolfe_PRD2011, eeQCD2017, Rougemont_arXiv2023, Critelli_PRD2017, Grefa_PRD2021, Grefa_PRD2022}. 
As sketched in Figure~\ref{fig:EDmodel}, the equation of state of this model contains a critical end point (full gray circle in Fig.~\ref{fig:EDmodel}(a)). The baryon density $\rho$ versus chemical potential $\mu$ at fixed $T $ has been calculated in Refs.~\cite{DeWolfe_PRD2011, Grefa_PRD2021} for different temperatures (see Fig.~\ref{fig:EDmodel}(b)). From $\rho(\mu)_T$, the quark susceptibility $\chi_2=d\rho/d\mu$ (proportional to the isothermal compressibility $\kappa_T = \rho^{-2} d\rho/d\mu$ in statistical physics) can be extracted (see Fig.~\ref{fig:EDmodel}(c)). Below $T_c$, $\rho(\mu)$ is multivalued as expected from a first-order transition. At $T_c$, $\rho(\mu)$ has an infinite slope at $\mu_c$, corresponding to a diverging quark susceptibility. Above $T_c$, $\rho(\mu)$ has an inflection point, which corresponds to a maximum of the quark susceptibility. The position of this maximum for different temperatures will delineate a line in the $T-\mu$ phase diagram, which is an often used proxy for the Widom line (see dotted red line in Fig.~\ref{fig:EDmodel}(a)). Refs.~\cite{eeQCD2017, Grefa_PRD2021} also calculated the thermodynamic entropy, which likely shows inflections at constant $T$ in the supercritical region, delineating another supercritical crossover. Ref.~\cite{Grefa_PRD2021} shows a crossover obtained from the locus of the minimum speed of sound squared, which lies close to the Widom line obtained from the quark susceptibility. 

A second example deals with a mapping of the QCD onto the three-dimensional (3D) Ising model, in the spirit of scaling theory~\cite{Schofield_1969, Schofield_Litster_Ho_1969, Guida:1997, ZinnJustin:2001}, as studied in many QCD works~\cite{Berdnikov_Rajagopal_2000, Nonaka:PRC2005, Stephanov_2011, Mukherjee_Venugopalan_Yin_2015, Monnai_Mukherjee_Yin_2017, An_2018, Parotto_PRC2020, Kahangirwe:arXiv2024}.
In this case, the Widom line is clearly expected to emerge from the critical end point.

Let us explain this in more detail. 
Close to the critical point $(T_c,\mu_{c})$, where $T_c$ is the critical temperature and $\mu_{c}$ is the critical baryon chemical potential, there is a linear invertible relation~\cite{Rehr_Mermin_1973,Luo_PRL2014} between physical variables $(T-T_c)/T_c$, $(\mu-\mu_{c})$ and 3D Ising model thermal field $t=(T-T_c)/T_c$ and magnetic field $H$. 
The equation of state is determined by this relation between physical variables and the universal equation of state of the 3D Ising model that depends on $t$ and $H$.  

The quantities $t$ and $H$ are parametrized as follows.
In the Ising model the free energy is well behaved along any contour leading from one side of the coexistence curve to the other side without crossing coexistence. 
As a result, Ref.~\cite{Schofield_1969} noted that close to the critical point it is possible to express the scaling fields in terms of the  variables $R$ and $\theta$. 
The former gives a measure of the distance from the critical point, and the latter measures the distance along a contour of constant $R$. 
With this transformation, the thermodynamic functions are well behaved in $\theta$ with the singular behavior controlled by $R$.

The parametrization of the QCD equation of state commonly used in QCD literature~\cite{Berdnikov_Rajagopal_2000, Nonaka:PRC2005, Parotto_PRC2020, Kahangirwe:arXiv2024} uses the linear mapping from physical variables to Ising variables, as mentioned above. 
Then, for the universal equation of state of the 3D Ising model, use is made of the so-called extended parametric model found in the work of Guida and Zinn-Justin~\cite{Guida:1997, ZinnJustin:2001}. 
With the notation of Ref.~\cite{ZinnJustin:2001}, the parametrization of the Ising variables $(t,H)$ in terms of $(R,\theta)$ is given by 
\begin{align}
    t&=R(1-\theta^2) \label{eq:r_QCD}\\
    H&=h_0 R^{\beta\delta} h(\theta) .
    \label{eq:tilde_h_QCD}
\end{align}
Here $h_0$ is a scale parameter, $\beta$ and $\delta$ are the usual critical exponents for the 3D Ising model, and $h(\theta)$ is a fifth-order polynomial of odd powers of $\theta$: $h(\theta) = \theta(1+a\theta^2 +b\theta^4)$. The best fit to available data obtained in Ref.~\cite{ZinnJustin:2001} gives $a=-0.76201$ and $ b=0.00804$, while $\epsilon$ expansion to order $\epsilon^3$ gives $ a=-0.72$ and $b=0.013$~\cite{ZinnJustin:2001}.
With this notation, the magnetization $M$, or more generally the order parameter, is given by
\begin{equation}
    M=m_0 R^\beta \theta , 
    \label{eq:M_QCD}
\end{equation}
i.e. is a linear function of $\theta$, with $m_0$ a scale parameter.
\begin{figure}[hb!]
\centering{
\includegraphics[width=0.999\linewidth]{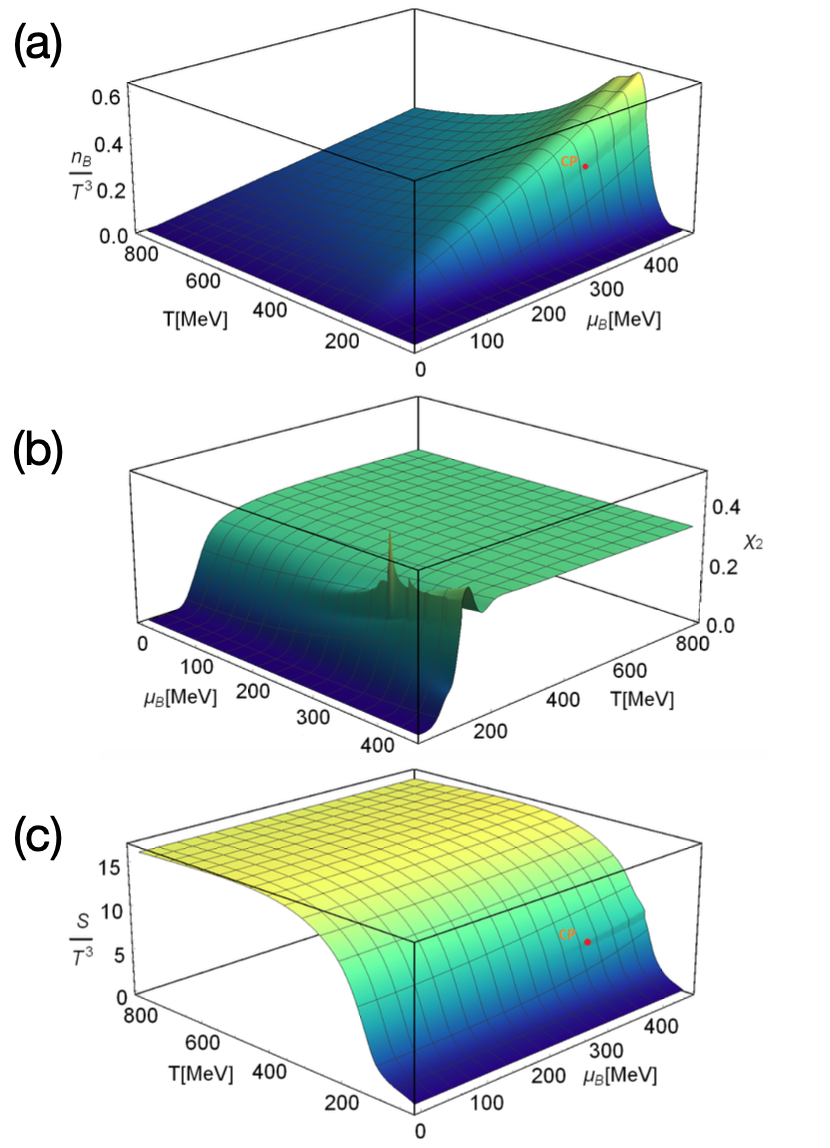}
}
\caption{QCD thermodynamics based on the mapping of the QCD onto the 3D Ising model obtained in Ref.~\cite{Parotto_PRC2020}. All panels are taken from Ref.~\cite{Parotto_PRC2020}. (a) Baryon density $n_B/T^3$ as a function of temperature $T$ and baryon chemical potential $\mu_B$. Red circle indicates the QCD critical end point, which in Ref.~\cite{Parotto_PRC2020} has been estimated to occur at $\mu_c=350$~MeV and $T_c\approx143.2$~MeV. Inflections develop in the supercritical region. (b) Second cumulant of the baryon number $\chi_2=T^{-2} \partial^2 P / \partial \mu_B^2$ as a function of $T$ and $\mu_B$. A peak appears and becomes sharper in magnitude upon approaching the critical end point. (c) Entropy density $S/T^3$ as a function of $T$ and $\mu_B$. Inflections occur in the supercritical region. 
}
\label{fig:Ising}
\end{figure}

A simpler approximation for the universal equation of state of the 3D Ising model was also studied in QCD~\cite{Stephanov_2011, Mukherjee_Venugopalan_Yin_2015, Monnai_Mukherjee_Yin_2017, An_2018}. It is based on the condensed-matter work of Schofield et al.~\cite{Schofield_Litster_Ho_1969} who studied CrBr$_3$, Ni, CO$_2$, Xe, He and $\beta-$brass above and below their critical points and found very good agreement between experiment and the simpler linear parametric model 
\begin{align}
    t&=R(1-b^2\theta^2) \label{r_Schofield} \\
    H&=h_0 R^{\beta\delta} \theta (1-\theta^2) \label{h_Schofield} \\
    M&=m_0 R^\beta \theta \label{M_Schofield} , 
\end{align}
with 
\begin{equation}
    b^2=\frac{\delta-3}{(\delta-1)(1-2\beta)} , 
\end{equation}
which is exact to order $\epsilon^2$~\cite{Brezin:PRL1972, Wallace:1974}.

The Widom line is the line of zero ordering field, and is thus given by $\theta=0$. On the other hand, the coexistence line is given by $\theta=\pm1$ in the linear parametric model of Schofield et al.~\cite{Schofield_Litster_Ho_1969}, whereas it is given by $\theta=\pm 1.154$ for the best nonlinear model in Eq.~\ref{eq:tilde_h_QCD}~\cite{ZinnJustin:2001}.

Indeed, as shown in Figure~\ref{fig:Ising}, taken from the recent Ref.~\cite{Parotto_PRC2020}, the baryon density (Fig.~\ref{fig:Ising}(a)) and the thermodynamic entropy  (Fig.~\ref{fig:Ising}(c)) all show inflections, which will delineate crossover lines in the supercritical region (see peak of $\chi_2$ in Fig.~\ref{fig:Ising}(b)).

Therefore our cursory survey shows that loci of extrema in different response functions are indeed displayed by simplified models of QCD whose equation of state contains a critical end point. Hence the Widom line is present in simplified models showing a critical end point in the QCD phase diagram. Rationalising the lines of inflections obtained in previous work as a ``Widom line" is the second main contribution of our work. Clearly, this result is expected, since it follows of course from the generality of the Widom line mechanism as an extension of the coexistence line in the supercritical region. Nevertheless, we believe that our work could open up future investigations to reanalyse some of these models within the framework of the Widom line.

\section{Significance of the Widom line in QCD}\label{sec:significance}
We have introduced the concept of Widom line for QCD matter and have shown that anomalous physical properties detected in previous theoretical work can be ascribed to Widom line phenomenology. The next step is to consider the significance or usefulness of the Widom line concept for QCD investigations. After all, as discussed in Sec.~\ref{Sec:models}, thermodynamic anomalies have already been detected and linked to the critical end point. Hence, from this perspective, the Widom line concept looks like just a new label for crossovers that have already been identified. 

As discussed in the introduction, the Widom line is an intrinsically interesting concept for describing supercritical region above a critical end point. Also, analogies with other fields often lead to cross fertilization of ideas that end up useful. Beyond these general considerations, and based on the analogies with work done in the context of supercooled fluids~\cite{water1, Franzese_JPCM2007} and doped cuprates~\cite{ssht, sshtSC, Lorenzo3band}, we suggest three directions where the Widom line has the potential to bring a novel perspective onto the QCD phase diagram. 

First, the Widom line can provide a unifying framework for describing and predicting the pronounced changes of thermodynamic anomalies at finite temperature and finite chemical potential in the QCD phase diagram. It rationalises existing crossovers into a single coherent theoretical framework, hence shedding new light on the origin of different crossovers, which can be attributed to the critical end point. 
As an analogy, the existence of anomalies in supercooled water was documented well before the introduction of the Widom line in 2005~\cite{water1}. Similarly, electronic phase crossovers in the hole-doped cuprates were documented well before the introduction of the Widom line in 2012~\cite{ssht}. In both fields, the Widom line emerged as a valuable concept to characterise the supercritical region. 

Second, the Widom line concept has predictive power. If a critical point exists, the anomalous behaviors in the phase diagram are attributed to fluctuations emanating from the critical end point. 
Hence, if the QCD critical point exists, then the Widom line can be used to predict its precise location in the phase diagram. This is because the Widom line concept contains a large amount of information about the supercritical state. The extrema of different thermodynamic response functions have defining features: upon approaching the critical end point, they become closer to each other, they narrow in width and they grow in magnitude, and eventually diverge at the critical end point. 
These characteristic signatures of the Widom line can be especially useful for QCD investigations, since the region at finite baryon chemical potential $\mu$ is difficult to access theoretically (e.g. sign problem in lattice QCD calculations). Ascertaining whether these distinctive signatures of the Widom line phenomenon (i.e. sharpening and narrowing of the extrema of response functions) occur as we move from $\mu=0$ to small values of $\mu$ in lattice QCD calculations or Taylor expansion methods can thus be used to infer what happens at large values $\mu$ and to estimate the location of the end point in the phase diagram. 

\begin{figure}
\centering{
\includegraphics[width=0.9\linewidth]{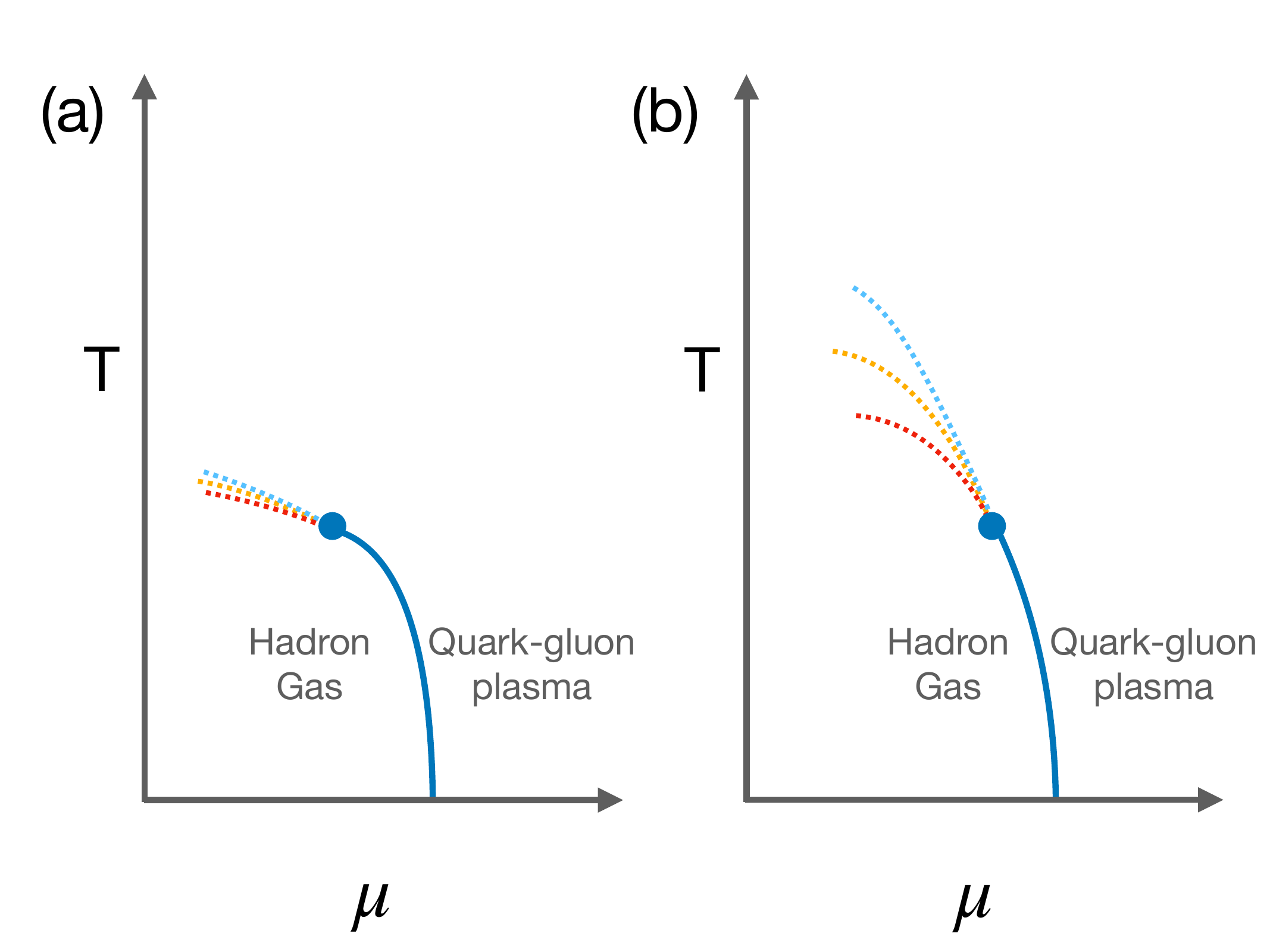}
}
\caption{Sketch of the QCD phase diagram for two different slopes of the first-order transition (solid line) close to the critical end point (full circle). Based on the results of Ref.~\cite{Luo_PRL2014}, the slope $dT/d\mu$ of the first-order transition close to the end point determines the speed of convergence of the loci of extrema of different response functions (indicated by three dotted lines) towards a single line  -- the Widom line -- asymptotically close to the end point. According to Ref.~\cite{Luo_PRL2014} the Widom line is usually found close to the loci of maximum isothermal compressibility. A small slope $|dT/d\mu|$ of the first-order transition close to the end point (panel (a)) implies that the crossovers in the different response functions remain close each other. On the other hand, a large slope $|dT/d\mu|$ of the first-order transition (panel (b)) implies that the crossovers in the different response functions separate from each other as we move away from the end point. The closeness in temperature of the crossovers at small chemical potential found in QCD calculations~\cite{BazavovPLB2019, Borsanyi:PRL2020} would imply a small slope of the first-order transition line, as sketched in panel (a). Charge-conjugation symmetry implies that the crossover lines must cross the $\mu=0$ line with zero slope.}
\label{fig:Convergence}
\end{figure}

Third, the Widom line is the continuation of the first-order transition in the supercritical region. This fact has two important consequences. (a) Close to the end point, the slope of the Widom line and of the first-order transition line are the same. Hence studying the Widom line will provide information on thermodynamic properties and stability of phases at the first-order transition between hadron gas and quark-gluon plasma. 
(b) More importantly, as shown in Ref.~\cite{Luo_PRL2014} using linear scaling theory, the slope of the first-order transition controls the order of the crossovers and determines the speed of convergence of the loci of extrema of different response functions towards a single line asymptotically close to the critical end point. The smaller the slope $dT/d\mu$ of the first-order transition in the $T-\mu$ QCD phase diagram, the closer the crossovers in the different response functions, as shown in Figure~\ref{fig:Convergence}.  
It is the non-universal relation between physical variables and the universal equation of state that leads to the different rates of convergence to the Widom line, as explained in Ref.~\cite{Luo_PRL2014}.

These insights could be useful for QCD investigations. The crossover at $\mu=0$ and $T\approx156.5$~MeV~\cite{BazavovPLB2019} from hadron gas to quark gluon plasma found in lattice QCD calculations has a narrow temperature width~\cite{BazavovPLB2019}. Studies with Taylor expansion method report that this crossover remains narrow as $\mu$ is increased, at least up to $\mu\approx 150$~MeV~\cite{BazavovPLB2019, Borsanyi:PRL2020}. Within the Widom line framework, the sharpness and closeness in temperature of the crossovers at zero and small chemical potential would imply a small slope of the first order transition line in the $T-\mu$ QCD phase diagram, in agreement with the findings of Refs.~\cite{Parotto_PRC2020, Kahangirwe:arXiv2024}. They would also imply that all crossovers lines will get closer each other upon increasing $\mu$. Furthermore, since response functions extrema converge asymptotically close to the critical end point, the temperature where they converge would give an upper boundary on the temperature of the critical end point, as pointed out in Refs.~\cite{water1, Franzese_JPCM2007}. 

\section{Conclusions}
In conclusion, we have introduced into the QCD phase diagram the concept of the Widom line, which is already widely used in the theory of critical phenomena of fluids and of correlated electron systems. In the QCD phase diagram, the Widom line, defined as the locus of maximum correlation length, is a supercritical crossover emanating from the critical end point of the hadron gas to quark-gluon plasma transition. 

We have surveyed some prior theory work based on simplified models of QCD whose equation of state contains a critical end point, to show that the thermodynamic anomalies that were identified emerging from the critical end point can be interpreted within the Widom line framework. 

Finally, we have suggested a few directions where the concept of Widom line could bring  new ideas on the QCD phase diagram: it can be a unifying framework for further theoretical exploration of the crossovers emerging from the end point, can be a useful tool to test the hypothesis of a critical end point in the QCD phase diagram and can link, through linear scaling theory,  the slope of the first-order transition to the speed of convergence of the loci of extrema of response functions towards a single line (the Widom line) asymptotically close to the critical end point.

The Widom line has a simple physical interpretation: in the equation of state of the 3D Ising model, it is parametrized by $\theta=0$, where $\theta$ measures the distance along a contour of constant $R$, so that thermodynamic functions are well behaved in $\theta$ with the singular behavior controlled by $R$. In the linear scaling theory framework, the Widom line is usually close to the loci of maximum isothermal compressibility~\cite{Luo_PRL2014}.

Further studies should examine the relevance of these theoretical suggestions. The discussion of the relevance of the Widom line concept for experimental results on heavy-ion collisions~\cite{StephanovRev, Stephanov:PRL1998, Stephanov:PRD1999, QCDAnnuRev2018} also deserves further investigations.

\begin{acknowledgments}
We thank Sanjay Reddy and especially Mikhail Stephanov for useful discussions and references. We also thank David S\'en\'echal for discussions and comments that improved the article. This work has been supported by the Canada First Research Excellence Fund. 
\end{acknowledgments}


%

\end{document}